\documentclass[a4paper,10pt,twoside]{article}

\usepackage{url}

\usepackage{graphicx}

\usepackage{tikz}

\usepackage{xcolor}

\begin{document}

\title{A NEW HOMEOSTATIC MODEL OF THE T CELL SYSTEM}

\author{TAM\'AS SZABADOS\footnote{Corresponding author, telephone: (+36 1) 463-1111/ext. 5907,
fax: (+36 1) 463-1677}\\
Department of Mathematics\\ 
Budapest University of Technology and Economics \\
M\H{u}egyetem rkp 3, Budapest, 1521, Hungary\\
\emph{email:} szabados@math.bme.hu \\
\emph{homepage:} www.math.bme.hu/\~{ }szabados \\ 
\\
TIBOR BAK\'ACS \\
Alfr\'ed R\'enyi Institute of Mathematics \\ 
Hungarian Academy of Sciences \\
Re\'altanoda u 13-15, Budapest, 1053, Hungary \\
\emph{email:} bakacs.tibor@upcmail.hu}

\maketitle

\begin{abstract}
Our main tenet argues that the primary role of positive thymic selection and the resulting T cell population is
the maintenance of a homeostatic equilibrium with self MHC-self peptide complexes. The homeostatic T cell repertoire can recognize infections non-specifically and this is an indirect (negative) recognition: the whole homeostatic T cell population together ``holds a mirror'' to the whole self, and any MHC-peptide complex that is ``not reflected in the mirror'' can be perceived by surrounding homeostatic T cells as a signal of the presence of a foreign entity. On the other hand, infection-specific T cell clones arise in a different pathway in the periphery, do not enter the thymus, and form a functionally different population. Here we summarize the basic assumptions and consequences of a logic-based new model, which differs from conventional models in many respects.
\end{abstract}



\section{Introduction}

\subsection{Fundamentals.}

The aim of this work is to describe a logic-based model of the fundamental operation of the T lymphocyte system. This means that the resulting model is obtained by a chain of logical arguments, though we are able to cite several supporting experimental references as well. We hope that our theoretical model can constitute a logically consistent basis to \emph{in silico} experiments with the immune system and can lead to new therapies as well.

Let us begin with some basic, more-or-less generally accepted assumptions, or even experimentally
established facts, about the human immune system that we need in this paper\cite{Jwbook,Paulbook}.
Normally, each human cell carries a large variety of different MHC I molecules on its surface. Their function is to present peptide fragments of inner proteins to the surveillance of the cytotoxic (CD8$^+$) T cell repertoire, which is a major constituent of the adaptive immune system. Ideally, the set of MHC I-peptide complexes on the surface of a cell represents accurately the inner metabolism of the cell, and, based on this information, the CD8$^+$ T cell repertoire is able to decide if there is a harmful entity inside or not.

The average length of a peptide attached to an MHC is about $10$ aminoacids, so there can potentially
exist about $20^{10} \approx 10^{13}$ different peptides. On the other hand, there can exist of the
order of $10^6$ ($\approx$ no. of different proteins $\times$ average length of a protein $/$ average
length of a peptide)  possible different self peptides, which is obviously a tiny fraction of the
potential variation. The number of different T cell clones at any moment is of the order of $10^6$ as
well, so the T cell repertoire should be carefully developed.

Thymus forms early in the embryonic development. Its main function is to develop and select a suitable
T cell repertoire from the ones randomly generated in the bone marrow. T cells with strongly
self-reactive receptors should be eliminated to prevent autoimmune reactions: this is called
\emph{negative selection}. It is much less obvious what \emph{positive selection} may mean. Since
genes encoding MHC proteins are highly polymorphic, there exists a large variation of individual MHC
proteins. Thus, it is clear that one function of the positive selection must be to select T cells that
are able to recognize self MHC molecules.

\subsection{What is the problem with conventional models of the T cell system?}

Standard textbooks in immunology teach that the role of positive selection is to identify and preserve
T cells that are likely to be able to respond to complexes of self MHC and \emph{foreign} peptides;
those that do not pass this test die due to neglect \cite{Jwbook,Paulbook}. A natural doubt may
arise then: How can the thymus of a fetus learn what would constitute a suitable T cell repertoire to
fight against an unknown universe of harmful agents, whose size is several magnitude larger than the
actual T cell repertoire of the fetus? How could the thymus biologically represent agents that the
fetus has never met? Since the choice of the foreign peptide repertoire should be very carefully made,
what inner force would make the choice and how?

On the other hand, even standard textbooks \cite{Jwbook} (Section III 7 Summary) mention that recognition of
self MHC-\emph{self} peptide complexes on thymic epithelial cells provides a positive survival
signal for developing T cells. It is referred to as \emph{``one of the central mysteries of
immunology''}.

Positive thymic selection has recently obtained much attention in the literature, cf. e.g. \cite{FOC00,KJHKC08,PZDRWC09,PN09,KHWK09}. It has become clear that T cells emerging from the thymus must react with
some self peptide-self MHC complex with not too strong affinity. One major consequence of medium self reactivity must be that thymus selected T cells create a homeostatic equilibrium with self cells in the periphery.

In the last decade there has been a growing number of evidence challenging the conventional view of thymus selected T cells, and emphasizing the homeostatic role of the T cell system, cf. e.g. \cite{PodTaubook,SSSYS01,SKB04,KBS04,MLC05,MS05,CZL07,HJ08,RRKCSS08,WZJ09,LFG09}. Our research group also described
some of the ideas of a homeostatic model in earlier papers \cite{BSVT01,BMSVT01,BMSVST07}, where
the reader can find more biological background as well. In the last cited paper we introduced the expression \emph{``homeostatic role of T cells model''}, or \emph{``HRT model''} for short, to refer to our approach.

We claim that if someone accepts that during embryonic development
the immune system of the fetus can only learn what constitutes his/her \emph{own self}, and
essentially nothing about future potentially harmful invaders, then practically every assumption in
this paper describing how the T cell system functions follows by logical necessity.

\section{Main Assumptions of a New Model} \label{sec:main}

\subsection{Positive selection in the thymus is based on self MHC-self peptide complexes.} \label{sec:possel}

Our main tenet argues that during embryonic development thymus forms to carry a nearly complete
repertoire of all self MHC-self peptide complexes that can be found in the fetus. In essence, the
immune system of the fetus learns what constitutes self. Negative selection eliminates dangerous T
lymphocytes that can bind too strongly to any self MHC-self peptide complexes and may cause autoimmunity. Positive selection supports T cells that can bind to some self MHC-self peptide complex with a medium, standardized
affinity. In this respect, the thymus works as a filter, letting only \emph{standardized T cells} out.

Fig. 1 shows a simplified graphical comparison between the new and the conventional models. Assume
that the shape of a \emph{T cell receptor (TCR)} is represented by a point of a so-called \emph{shape
space}. For visualization purposes, it can be assumed that the space of shapes and other chemical and
physical properties relevant for binding between a ligand and a receptor can be represented by a large
region in the plane. Only complementary or nearly complementary shaped ligands and receptors can bind.
Similarity of shapes corresponds to metric distance in our simplified scheme; that is, the affinity
depends on the distance of the corresponding points in the shape space. The figure shows a tiny
subrectangle of the shape space. The dots in the figure represent TCR's that are exactly complementary
to some self MHC-self peptide complex. The areas shaded in gray represent the set of shapes that are
allocated to possible TCR's after negative and positive selection in the two respective models.

\begin{center}

\begin{tikzpicture}[scale=0.8]
  \draw (0,0) rectangle +(5,-7);
  \node[below] at (2.5,-7) {\small{New model}};
  \filldraw[fill=gray!60] [even odd rule] (1,-1) circle (0.5) circle (0.25);
  \fill (1,-1) circle (0.06);
  \filldraw[fill=gray!60] [even odd rule] (3,-4) circle (0.5) circle (0.25);
  \fill (3,-4) circle (0.06);
  \filldraw[fill=gray!60] [even odd rule] (4.2,-2) circle (0.5) circle (0.25);
  \fill (4.2,-2) circle (0.06);
  \filldraw[fill=gray!60] [even odd rule] (4.1,-6.2) circle (0.5) circle (0.25);
  \fill (4.1,-6.2) circle (0.06);

  \node[below] at (9.5,-7) {\small{Conventional models}};
  \filldraw[fill=gray!60, even odd rule] (7,0) rectangle +(5,-7);
  \filldraw[fill=white] (8,-1) circle (0.5);
  \fill (8,-1) circle (0.06);
  \filldraw[fill=white] (10,-4) circle (0.5);
  \fill (10,-4) circle (0.06);
  \filldraw[fill=white] (11.2,-2) circle (0.5);
  \fill (11.2,-2) circle (0.06);
  \filldraw[fill=white] (11.1,-6.2) circle (0.5);
  \fill (11.1,-6.2) circle (0.06);

\end{tikzpicture}

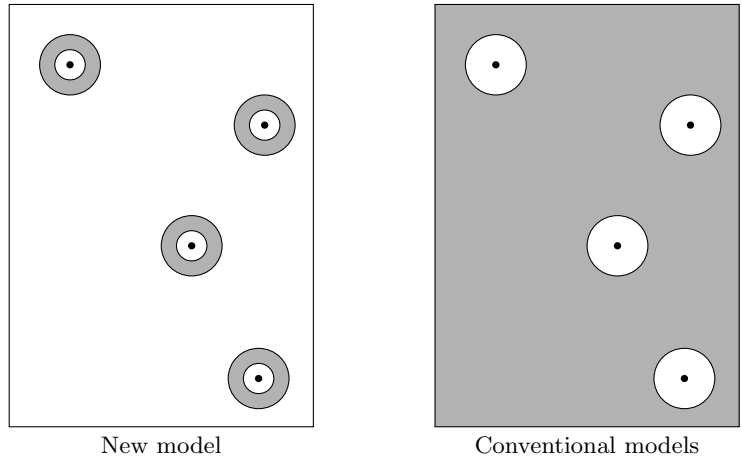
\small{Fig. 1.  Graphical representation of the difference between the new homeostatic and the conventional
models}

\end{center}

The figure shows that according to conventional models a large and supposedly unknown gray region
should be covered by the T cell repertoire by birth. In that model the gray region is the set of all
possible TCR shapes excluding the complementary neighborhoods of self MHC-\emph{self} peptide complexes.

On the other hand, according to our new model much smaller gray rings show possible positions of the
elements of the T cell repertoire: ideally, at least one T cell clone (one kind of TCR) is centered in
each gray ring. Moreover, these gray rings can be completely known and represented by the thymus by
the time of birth, since they are uniquely determined by the self cells. This means that in our model
the T cell repertoire of a fetus (that has not met infection yet) consists of T cells that are able to
bind some self MHC-self peptide complex with medium avidity, but cannot bind any self MHC-self peptide as strongly as to start a dangerous autoimmune reaction. That is why the grey rings of our model cover the outer part of the white discs of a conventional model.

\subsection{The thymus selected T cell repertoire has a homeostatic role.} \label{sec:homeo}

Accepting the fact that for the survival of a T cell clone it is necessary to have a regular stimulus from binding MHC-peptide complexes \cite{FR00,KBB97}, it follows that in a healthy state of the individual there must be a dynamic equilibrium between the T cell repertoire and the repertoire of different MHC-self peptide complexes. This results in a thymus-selected \emph{homeostatic T cell repertoire} which establishes a \emph{homeostatic control} for both self cells and T cells. This control sets upper and lower limits for the number of cells in each specific T cell clone and also for the number of corresponding self cells at any given time. While in normal, healthy circumstances tissues are not penetrable for T cells, they are penetrable for TCR's shed by T cells. Tissue cells also regularly shed their MHC I-peptide complexes into the circulating lymph. Thus, extending the previously mentioned homeostatic network with soluble TCR's and MHC-peptide complexes, we assume that the body has a practically complete surveillance system, which takes care of inner tissue cells as well and is in equilibrium in a healthy individual.

This homeostatic control can explain why the role of thymus decreases with age, and negative and
positive selection in the thymus is gradually replaced by the homeostatic control of the periphery. In
fact, a too weakly binding T cell clone does not get suitable stimulus and ``dies by neglect''; while a
too strongly binding T cell clone has not enough time to divide. Thus a medium avidity is optimal for
the survival of a T cell clone. This observation is consistent with the \emph{molecular
complementarity principle} \cite{RD97,DR97}.

\subsection{Recognition of non-self entities or mutated self cells is carried out
by the whole homeostatic T cell repertoire.} \label{sec:whole}

In this section we only consider infections or mutations for which there exists no immune memory;
creation of infection-specific clones are discussed in Section \ref{sec:infspec}.

In contrast to conventional models, one of our major tenets says that a single T cell cannot and should not decide alone what is self and what is non-self. Shapes of self and non-self entities are intricately interwoven sets; in the language of the shape space model, the subsets of points representing self and non-self are complexly interlaced and cannot be separated by a nice smooth mathematical curve that even an individual T cell might ``memorize'': the complexity of the antigen universe exceeds the capacity of an individual T cell. The ``knowledge'' of each specific T cell is represented by the shapes of its receptors, especially by its TCR. This implies that the task an individual T cell can carry out is to recognize complementary or near complementary MHC-peptide complexes. This way a specific T cell clone is responsible for the population of complementarily shaped specific self MHC-self peptide complexes in the body.

Only the whole thymus-selected T cell population is able to answer the fundamental question of immune recognition, self/non-self discrimination. Although homeostatic T cells act independently of each other, together they ``hold a mirror'' to the whole self; their whole repertoire accurately reflects the whole self antigen population, and this can be used to distinguish self and non-self in the way described below. Briefly, a peptide belongs to non-self it is ``not reflected in the mirror''.

A tissue cell carries a large variety of \emph{different} MHC I-peptide complexes. If the cell is
healthy, then each of these specific complexes are regularly bound by their own specific complementary
TCR's circulating in the lymph in the neighborhood of the cell. This is a vital feedback to the cell
that its metabolism works all right.

After a new viral infection or mutation affecting the cell's metabolism, new non-self or new mutated self peptides appear in some of its surface MHC I molecules. We assume that the typical consequence of this is that for \emph{a critically long period of time} these MHC I-foreign peptide complexes are \emph{not} bound by \emph{any} surrounding TCR's. This way, the cell receives a vital signal from \emph{the whole homeostatic T cell system}: ``something is wrong inside''. Again, by simple logic, in this multiplayer game in the neighborhood of the infected cell the only entity that can ``know of'' the presence of e.g. a viral invader is the infected cell itself, after not getting a binding partner for some of its surface MHC I molecule for a critically long period of time. Thus it is reasonable to assume that such an infected cell becomes \emph{excited}, ready to signal the problem to its environment. One important aspect of our new homeostatic model is that the recognition of a dangerous new agent is an \emph{indirect (negative) recognition}: the absence of existing TCR partners signals the problem.

Other elements of the immune system, especially T cells, can get the information that there is a
problem in two ways. One way is that \emph{alarm signals}, that is, specific chemokines, are secreted
by an infected cell into its environment. Their role will be explained in paragraph \ref{sec:alarm}; however,
there is another pathway through lymph nodes as well, described in the next paragraph.

An important advantage of the non-specific, polyclonal reaction of the ready and always present homeostatic T cell repertoire against a \emph{new} foreign entity is that they can react much faster than an infection-specific T cell clone of the conventional models: the latter's creation can last too long to stop e.g. a rapidly replicating virus.

\subsection{An alarm signal arises in a lymph node if there is an invader in the periphery.} \label{sec:lymph}

It is experimentally observed that somatic cells shed their MHC I-peptide complexes into the lymph. Similarly as with the thymus, we assume that in a healthy individual there exists a practically complete homeostatic repertoire of CD8$^+$ T cells and TCR molecules in a lymph node to capture almost all soluble self MHC I-\emph{self} peptide complexes. On the other hand, MHC I-\emph{foreign} peptide complexes cannot be captured by any TCR's present in a lymph node, so the concentration of the latter complexes will be larger at certain parts of the lymph node, where they can be captured by dendritic APC's. This filtering effect is a kind of reverse of the filtering in the thymus: the thymus lets through T cells that are able to interact with some self MHC-self peptide complexes (with standardized, medium strength affinity, cf. \ref{sec:possel}), while pathways in a lymph node let through only those soluble MHC-peptide complexes that are foreign (not complementary) to the homeostatic T cell repertoire.

In general, APC's present peptide fragments of native proteins sampled from the humoral phase to the
surveillance of circulating helper (CD4$^+$) T cells. This presentation occurs on surface MHC II molecules of the
APC's. In a healthy individual, a dendritic APC in a lymph node
will present only self peptide complexes on its MHC II molecules. This means that all of the
MHC II-peptide complexes of the dendritic cell will be regularly visited by specific complementary
cells of the homeostatic CD4$^+$ T cell repertoire. However, after infection or mutation, some of the surface
MHC II molecules on some dendritic cells will present foreign peptides. Consequently these new
complexes \emph{will not be bound by any member of the homeostatic helper T cell repertoire for a
critical period of time}. This is a vital signal for a sentinel dendritic cell: it shows that there is a
problem somewhere in the periphery. Thus this dendritic cell gets \emph{excited} and releases \emph{alarm
signals}: specific interleukins to helper and cytotoxic T cells in the lymph node. As
a result, a non-specific, polyclonal reaction follows: many of the T cells leave the lymph
node to search for invaders in the periphery. Again, it should be emphasized that -- in contrast to conventional models -- not individual T cells, but only the whole homeostatic repertoire of CD4$^+$ T cells can recognize the presence of a foreign invader.

\subsection{CD8$^+$ T cells find their targets in the periphery by the help of alarm signals.} \label{sec:alarm}

How can activated T cells find their targets in the periphery? As it was described above, after a while infected or mutated tissue cells start secreting \emph{alarm signals}, that is, specific chemokines. These signals help cytotoxic T cells to penetrate the corresponding tissue and to find their targets to destroy. We call this the \emph{``smoking gun''} phenomenon. We assume that such an alarm signal leads to a \emph{non-specific, polyclonal} immune answer from any CD8$^+$ T cell getting close to affected tissue cells, lysing infected cells, that is, cells that are in an excited state. We refer to this phenomenon as physiological transplantation reaction.

In fact, the strong reaction of the T cell system against transplanted tissues cannot be found under natural conditions, so it needs a physiological explanation. Our model implies that this strong reaction is ignited by alarm signals secreted by all excited cells in a transplanted tissue because they all are missing their complementary TCR partners.

We mention that alarm signals in our model significantly differ from the danger signals in Matzinger's theory, where danger signals alert the immune system to \emph{dying cells} \cite{GM01,SER03}.

The above described main assumptions can hopefully explain why we call our model \emph{``homeostatic role of T cells model''} \emph{(HRT model)}.

\section{Infection-Specific T Cell Clones in the HRT Model} \label{sec:infspec}

It is important that our HRT model can and does accommodate infection-specific T cell clones as well.
The ideas described in Section \ref{sec:main} correspond to the case when the immune system meets a \emph{new} infecting agent; moreover, they describe the behavior of the T cells in the \emph{initial phase} of a new infection.

Immature T cells are constantly and randomly generated in the bone marrow and they circulate before
the thymus filters out the ones unsuitable for the homeostatic surveillance. If a new infection
persists for several days, then a few specific T cells may randomly be generated with high affinity to peptides
of the infection presented by MHC's. By chance, even a thymus-standardized T cell can have high
affinity to a new infection. After a while, in the periphery, T cells of either origin may create a new specific clone against the infection and support immune memory.

How can the immune system distinguish T cells that are part of the homeostatic system and T cells with
high affinity to a foreign peptide? Obviously, this distinction is very important. A thymus filtered T
cell that is part of the homeostatic equilibrium can attach to specific self MHC-self peptide
complexes with a neither too strong, nor too weak, standardized  affinity. When a tissue cell is
healthy, then all of its surface MHC I carries a self peptide, and in this case all of these complexes
are regularly visited by their respective specific TCR's that attach with standardized affinity. For
brevity, thymus-standardized T cells participating in the homeostasis are called \emph{homeostatic
T cells} in this paper.

On the other hand, a CD8$^+$ T cell whose receptor has high affinity to a self MHC I - foreign peptide
complex (for brevity: \emph{an infection-specific CD8$^+$ T cell}) can reach an infected tissue cell only if each of the following conditions holds:

\begin{itemize}

\item
The infected tissue cell has not received complementary homeostatic T cell visitors to some of its MHC I's that present foreign peptides for a critically long period of time. Thus this cell has become excited, and is releasing alarm signals in the form of chemokines. This supposedly has an effect on the behavior of any approaching T cell, signalling the presence of a foreign entity.

\item
The contact between an infection-specific T cell and an infected tissue cell takes place at an
\emph{exciting MHC-foreign peptide complex}, that is, at a complex which causes the excitation
of the infected cell. We suppose that this circumstance has a biochemical influence on the behavior of
both contacting partners. It must be assumed that such a special contact does not have a homeostatic,
``soothing'' effect on the infected cell, eliminating its excitation, since in that case the cell
would lose the vital information that it is infected.

\item The affinity of the receptor of an infection-specific T cell with the relevant MHC-foreign
peptide complex should be significantly larger than the thymus standard intermediate affinity of the
homeostatic T cells.

\end{itemize}

This way, an infection-specific cytotoxic T cell contacting an infected cell can get the information
that this attachment is not of the homeostatic kind, but a meeting with an infected cell. As a result,
it can lyse the infected cell and can proliferate and differentiate into effector cells capable of
combating the infection. Thus, eventually an infection-specific T cell clone is created. This
process is supported by helper T cells as well: this follows from the explanation in the next section.

But if there exists a non-thymic pathway to create a T cell clone, and so some T cells do not undergo negative selection, they may recognize purely self antigens, which in principle may lead to autoimmunity. However, the HRT model is protected against such danger as described by the conditions above. Namely, if a randomly generated CD8$^+$ T cell that has not gone through the thymus recognizes a \emph{self} peptide-MHC complex, then this complex cannot be in an exciting state, because each self peptide-MHC complex is regularly visited by complementary homeostatic T cells by the axioms of the HRT model. (Remember that excitation is caused by the absence of visiting complementary homeostatic T cells.) Thus, in general, the carrying cell is not in excited state either and so does not secrete alarm signals. Hence the contacting CD8$^+$ T cell does not get the necessary biochemical signals (so to say, does not get permission) to lyse the contacted cell, or to proliferate, or to differentiate into an effector cell. So in healthy circumstances, such a dangerous T cell cannot create a T cell clone.

To create immune memory of an infection, that is, an infection-specific T cell clone of significant
longevity, plus functionally distinguish the infection-specific T cell repertoire from the homeostatic T
cell repertoire, the following assumptions seem essential as well:
\begin{itemize}
\item The thymus does not support the creation of an infection-specific T cell clone, nor does it
maintain its existence; this clone should be kept alive in the periphery.

\item Lymph node filters, described in \ref{sec:lymph}, do not filter out MHC I-infection-specific peptide
complexes; in contrast, they capture only MHC I-self peptide complexes. So these MHC I-infection-specific peptide complexes can be taken up by APC's in the lymph node.

\item Some of the APC's in lymph nodes or in the periphery should be constantly in a state of
excitation, carrying exciting MHC-infection-specific peptide complexes, to keep an infection-specific T cell clone alive. In other words, regular stimulation from APC's is needed -- even when the infection has been overcome -- to maintain the existence of an infection-specific T cell clone \cite{ACT09}.
\end{itemize}
If these conditions do not hold, then the immune memory to this specific infection is lost.

\section{Extension of the HRT Model to Humoral Immune Reaction and Helper T Cells}

A mechanism closely related to the one described above can explain the humoral immune response as
well. In a healthy individual, B cells meet only self (or other harmless) native soluble
antigens in the serum. Those B cells that can attach to any of these self antigens present only self peptides on
their surface MHC II molecules. Similar is the case with other APC's. Consequently, thymus selected
helper T cells can meet only self MHC II-self peptide complexes in the serum. Since they
attach with standardized medium affinity to their specific targets, they send a signal to B cells:
``everything is all right in the serum''. This is part of the homeostatic control described
above. It is important in our new model that specific CD4$^+$ T cell clones should exist for
practically all kinds of self MHC II-self peptide complexes normally present in any of the APC's.

A new (e.g. bacterial) infection leads to the appearance of new native antigens in the serum.
The first line of defence is the innate immune system that normally begins the fight against the new
invader. Soon some of the APC's, in particular B cells with some affinity to the new antigens, will
present new, foreign peptides on their surface MHC II proteins. Then typically \emph{these new
complexes will not be bound by any helper T cell of the homeostatic repertoire for a critically long
time}. This way the carrying B cell gets the information that some of its surface MHC II has a peptide
of an invader. This B cell gets into an \emph{excited} state and starts to release suitable
\emph{alarm signals}, that is, suitable interleukins.

This signal initiates a non-specific, polyclonal activation in surrounding T helper lymphocytes. As a
response, they release interleukins, which stimulate the \emph{excited} B cells to division,
hypermutation, and eventually to become plasma cells. Clearly, this process may require several days;
during this period the innate immune system, including its APC's that also obtain the interleukin
help, has to defend the host against the circulating infecting agents. Activated cytotoxic T cells, in
particular the ones specific to the infection, can also get helper interleukins that support their
division. Again, we emphasize that by the HRT model -- unlike with conventional models -- \emph{only the whole homeostatic repertoire of CD4$^+$ T cells can distinguish self and non-self}: one cannot load this burden on individual T cells.

This process gets more impetus when activated T cells leave lymph nodes and appear in the humoral
phase, as described in the previous section. Under favorable circumstances, after a longer period of
time, new infection-specific helper T cell clones can arise as well, by a similar process as the one
described in the previous section for infection-specific cytotoxic T cells.

A brief comparison of the main assumptions and implications of our HRT model and the conventional models is included in  Table \ref{table}.

\begin{table}
  \centering

\begin{tabular}{|p{40mm}|p{40mm}|p{40mm}|}
  \hline

 & \begin{center}\textbf{HRT model} \end{center} & \begin{center}\textbf{Conventional models\cite{Jwbook,Paulbook}}\end{center} \\
  \hline
 \emph{Positive selection supports T cells}  & that have medium, standardized self-reactivity; &  that are likely
 to be able to respond to foreign peptides. \\
 \emph{The thymus selected T cell repertoire}  & has a homeostatic role; & does not react with self. \\
 \emph{Primary recognition of non-self is carried out } & by the whole homeostatic T cell repertoire in an indirect way; & by individual T cells in a direct way.\\
 \emph{Infection-specific T cells can recognize non-self} & only with the help of the whole homeostatic T cell repertoire; & on their own. \\
  \emph{Infection-specific T cell clones are created} &  in a non-thymic pathway in the periphery; & in a thymic pathway. \\

  \hline
\end{tabular}

  \caption{Comparison of the HRT and conventional models}\label{table}

\end{table}

\section{Experiments Verifying the Assumptions of the HRT Model}

The HRT model predicts that in a fetus the T cell repertoire reflects only self antigens. That is, any
T cell or TCR from such a repertoire can interact with some self MHC-self peptide, with some medium
affinity. Several experiments \cite{CKFC01,LBL02,MSP02,MMS03,SZL03,SHA04} seem to support that T cells in newborns are of homeostatic type. Also, an important experiment \cite{AHPT02} shows the presence of lymphocyte reaction to self antigens in patients with no autoimmune diseases. Another experiment \cite{SMWD08} verifies that T cells play an important role in tissue regeneration.

Recently, Wang et al.\cite{WPM09} generated transgenic (3KO) mice that express a single peptide-MHC class I
complex engineered as single chain trimers to demonstrate ``that positive selection of CD8$^+$ T
cells in these mice results in an MHC-specific repertoire''. These T cells are highly MHC and peptide
specific, governed by positive selection based on self-recognition. Thus, these T cells would seem to
be ideal for testing some key predictions of our model.

\section{Potential Applications Based on the HRT Model}

\subsection{Potential application in transplantation.}

Acquired immune tolerance -- an experimental manipulation by which the content of self-components is
enlarged -- means that the allograft is accepted without immunosuppression while preserving the
recipient's protective immunity. It represents a solution to the problems of acute and chronic
rejection and the resulting long-term reliance on toxic immunosuppressive therapies. Although some
historical clinical observations showed clues about how to induce tolerance without the aid of
immunosuppression in organ transplantation as early as the 1960s, efforts for its exploitation did not
start for more than half a century\cite{SMDTAG04}.

One idea to achieve acquired immune tolerance that is consistent with the HRT model is as follows. If
a transplantation is escorted by a transplantation of the donor's lymphocytes or soluble lymphocyte
receptors as well, then that can lead to a \emph{microchimerism} in the transplanted organ. The aim is
that MHC I-peptide complexes of donor tissue cells be constantly visited by the necessary repertoire
of donor TCR's. By the HRT model, this inhibits donor tissue cells from releasing alarm signals, so
the recipient's cytotoxic T cells will not attack them. If donor T cells and TCR's can be made
constantly present in the recipient's neighboring lymph nodes as well, the HRT model predicts even
better results. This is due to the fact that then soluble donor MHC-peptide complexes will be filtered out in these lymph nodes, so they cannot initiate the appearance of activated T cells in the transplanted tissue. 

In the literature, supporting evidence for this claim is growing. For example, an important experiment\cite{KMKEF03} describes a so-called ``infectious tolerance'' in rat or mouse heart, skin, or liver transplants: ``Regulatory cells developing after the spontaneous acceptance of a Lewis to DA liver transplant can serially adoptively transfer the acceptance of a Lewis rat cardiac graft in spite of the presence of in vitro antidonor reactivity. Both CD4$^+$ and CD8$^+$ populations have this regulatory activity.'' Another experiment\cite{WAFC08} proves that recombinant TCR ligands can be used to treat autoimmune diseases. A recent observation\cite{KB10} verifies that regulatory T cells contribute to maternal tolerance of the fetus. These observations seem to support the hypothesis that T cell chimerism may contribute to tolerance.

\subsection{Potential application in cancer treatment.}

A successful tumor cell must avoid or at least decrease the attacks of the immune system on it\cite{K09}. One
possibility is that the tumor cell avoids or significantly decreases the presentation of antigens on
its surface MHC I molecules. Another possibility is that a cancer cell expresses only normal
self-peptides on its surface MHC I molecules. A third possibility is that tumor cells may extend the
homeostatic T cell repertoire by recruiting complementary T cells from naive T cells that are
constantly produced by the bone marrow (e.g. during long lasting chemical carcinogenesis such as
smoking). In such a way, a successful tumor cell may create a tumor-specific homeostatic-like environment
in its neighborhood or in lymph nodes, similarly to the microchimerism in the case of
transplantation, described above.

It is a well-known fact that, as a result, most tumor cells have no immunogenic effects. Then the HRT
model implies that tumor cells are, paradoxically, under homeostatic T cell control as well. This
seems to conform with experimental observations\cite{DRHBG08}. According to our HRT hypothesis, resection of the primary tumor mass is perceived as loss of 'normal' tissue cells. Consequently, T cells striving to reconstitute homeostasis stimulate rather than inhibit the growth of dormant tumor cells and avascular micrometastases. We therefore suggested that such kick-start growths could be prevented by a recombinant T cell receptor ligand therapy that modifies T cell behavior through a partial activation mechanism\cite{BM10}.

\section{Conclusions}

A new, logic-based model of the fundamental operation of the T cell system has been described above. We think that it can better explain some rather poorly understood phenomena than conventional models. The most essential novelties are: a more logical explanation of positive thymic selection; the assumed presence of an always ready homeostatic T cell repertoire for fast response to new infections or mutations; indirect (negative) recognition of non-self antigens by the whole homeostatic T cell repertoire; a non-thymic pathway to the creation of infection-specific T cell clones; a somewhat novel explanation of transplantation reaction and the suggestion of a new, non-immunosuppressive therapy; and, finally, the suggestion of a therapy to avoid stimulation of dormant tumor cells after the resection of the primary tumor.


\section*{Acknowledgments}

The authors are very much grateful to G\'abor Tusn\'ady, P\'eter L. Erd\H{o}s, and L\'aszl\'o \'Ury
(Alfr\'ed R\'enyi Institute of Mathematics, Budapest) for many useful discussions of the ideas
underlying the present paper during the past years.  Also, we are obliged to Professor Jay Mehrishi
(Cambridge University), Professor Michael Steinitz (The Hebrew University, Jerusalem), and an anonymous referee for their valuable remarks. T.S. thanks his student Krist\'of H\"or\"omp\"oly for creating Fig.1 and for some useful ideas.


\section*{References}

\bibliography{new_model_of_T_cells_arxive}

\bibliographystyle{unsrt}

\end{document}